\documentclass{aastex62}
\usepackage{natbib}
\usepackage{graphicx}
\usepackage{multirow}
\usepackage{amsmath}	% Advanced maths commands
\usepackage{amssymb}	% Extra maths symbols
\usepackage{bm}		% Bold maths symbols, including upright Greek
\usepackage{color}
\usepackage{float}
\usepackage{lineno}
\usepackage[figuresright]{rotating}

\received{-}
\revised{-}
\accepted{-}
\submitjournal{ApJ}

\shorttitle{Searching for the TeV candidates in 4LAC HBLs}
\shortauthors{Zhu et al.}

\begin{document}
%\linenumbers

\title{Searching for TeV Candidates in 4LAC High-synchrotron-peaked Frequency BL Lac Objects}

\correspondingauthor{Y.G. Zheng}\email{ynzyg@ynu.edu.cn}

\author{K.R.Zhu}
\affiliation{Department of Physics, Yunnan Normal University, Kunming, Yunnan, 650092, People's Republic of China}

\author[0000-0002-9071-5469]{S.J. Kang}
\affiliation{School of Physics and Electrical Engineering, Liupanshui Normal University, Liupanshui, Guizhou, 553004, People's Republic of China}

\author{R.X.Zhou}
\affiliation{Department of Physics, Yunnan Normal University, Kunming, Yunnan, 650092, People's Republic of China}

\author{Y.G. Zheng}
\affiliation{Department of Physics, Yunnan Normal University, Kunming, Yunnan, 650092, People's Republic of China}

\begin{abstract}

The next generation of TeV detectors is expected to have a significantly enhanced performance. It is therefore constructive to search for new TeV candidates for observation. This paper focuses on TeV candidates among the high-synchrotron-peaked BL Lacertae objects (HBLs) reported in the fourth catalog of active galactic nuclei detected by the Fermi's Large Area Telescope, i.e., 4LAC. By cross-matching the Fermi data with radio and optical observations, we collected the multiwavelength features of 180 HBLs with known redshift. The data set contains 39 confirmed TeV sources and 141 objects whose TeV detection has not yet been reported (either not yet observed, or observed but not detected). Using two kinds of supervised machine-learning (SML) methods, we searched for new possible TeV candidates (PTCs) among the nondetected objects by assessing the similarity of their multi-wavelength properties to existing TeV-detected objects. The classification results of the two SML classifiers were combined and the 24 highest-confidence PTCs were proposed as the best candidates. We calculate, here, the 12 year averaged Fermi spectra of these PTCs and estimate their detectability by extrapolating the Fermi spectrum and including the extragalactic background light attenuation. Four candidates are suggested to have a high likelihood of being detected by the Large High Altitude Air Shower Observatory and 24 are candidates for the Cerenkov Telescope Array observations.

\end{abstract}

\section{Introduction}\label{sec:intro}
Most extragalactic sources detected in the $\gamma$-ray band belong to the blazar category\citep{2020ApJS..247...33A}. Blazars are an important subclass of active galactic nuclei (AGNs) and are characterized by their strong and rapid variability and high levels of brightness (e.g., \citealt{1978PhyS...17..265B,1995PASP..107..803U}). The spectral energy distributions (SEDs) of blazars are dominated by two components, which are illustrated by a double-bump spectral shape in $\rm log\nu$-$\rm log\nu F\nu$ space. The origin of the low energy bump, seen from the radio band to the ultraviolet or soft X-ray band, is attributed to the synchrotron emission of a relativistic electrons in the jet.  Either leptonic models (e.g., \citealt{1992A&A...256L..27D,1992ApJ...397L...5M,1993ApJ...416..458D,1996ApJ...461..657B,2016MNRAS.457.3535Z}) or hadronic models (e.g., \citealt{2000NewA....5..377A,2001ICRC....3.1153M,2003APh....18..593M,2013ApJ...764..113Z}) can be used to reproduce the high energy emission of blazars. According to the presence or absence of broad emission lines in their optical spectra, blazars are divided into flat spectrum radio quasars (FSRQs) and BL Lac objects. The equivalent widths (EWs) of FSRQ optical spectra emission lines in the comoving frame are greater than 5$\rm \AA$, while the EWs of BL Lac objects are less than 5$\rm \AA$ \citep{1991ApJ...374..431S}. The peak frequency, $\nu_{\rm syn}$, of the low energy bump (synchrotron bump) can also be used to classify blazars, as follows: low-synchrotron-peak blazars (LSP; $10^{14}\leq \nu_{\rm syn}\leq 10^{15} \rm{Hz}$); intermediate-synchrotron-peak blazars (ISP; $10^{14}\leq \nu_{\rm syn}\leq 10^{15} \rm{Hz}$); high-synchrotron-peak blazars (HSP; $\nu_{\rm syn}\textgreater 10^{15} \rm{Hz}$) \citep{2010ApJ...716...30A}; and extreme HSP blazars (EHSP, $\nu_{\rm syn}\textgreater 10^{17} \rm{Hz}$) \citep{2018MNRAS.480.2165A}. Multiwavelength observations show that the origin of the nonthermal emission of blazars extends from the radio band to the $\gamma$-ray band. Some can even extend to the very high energy band (VHE, $E \textgreater 0.1\ \rm{TeV}$, e.g., \citealt{2008sf2a.conf..297D}).

VHE $\gamma$-ray astronomy research focuses on the high energy emissions from extreme objects and their physical mechanisms. Research regarding TeV sources has promoted the development of both VHE $\gamma$-ray astronomy and neutrino astronomy \citep{2015A&A...579A..34A,2017A&A...598A..17C, 2019A&A...632A..77C,2019ApJ...887..104C}. However, this research has long been hindered by limited observations across the TeV energy band and the limited number of TeV sources. Currently, the observation of TeV sources mainly depends on imaging atmospheric Cerenkov telescopes (IACTs), such as WHIPPLE, MAGIC, H.E.S.S., and VERITAS, or extensive air shower (EAS) arrays, such as Tibet AS-$\gamma$ and ARGO-YBJ. Traditional ground-based detectors suffer from a small field of view (FOV), limited sensitivity, and short exposure times. They are also affected by weather and the Earth's magnetic field. In addition, although a number of sources emit TeV $\gamma$-rays, it is difficult for TeV photons to reach the Earth because of the absorption caused by photons from extragalactic background light (EBL) or cosmic microwave background (CMB), with which they interact. This is especially the case for distant sources \citep{2001ARA&A..39..249H, 2011MNRAS.410.2556D, 2018A&A...614C...1F}. Taking these various issues together, TeV sources are difficult to detect and identify. The online VHE $\gamma$-Ray Source Catalog, TeVCat\footnote{TeVCat \url{http://tevcat.uchicago.edu/}} \citep{2008HEAD...10.4106H} reports 232 TeV sources so far. There are 77 TeV blazars, including 66 BL Lacs, which dominate the extragalactic TeV sources. The next generation of TeV detectors, i.e., the Cherenkov Telescope Array (CTA) and the Large High Altitude Air Shower Observatory (LHAASO), are expected to perform significantly better. This makes it worthwhile to search for a collection of high-confidence TeV candidates for follow-up observations.

The Fermi Large Area Telescope (Fermi-LAT) has observed the entire sky at GeV energy levels for more than 12 years \citep{2009ApJ...697.1071A}. Making use of these Fermi GeV $\gamma$-ray observations provides an alternative approach to explore for TeV candidates. On the basis of the first seven years of observations, the Third Catalog of Hard Fermi-LAT Sources (3FHL, \citealt{2017ApJS..232...18A}) reported 1556 sources detected in the energy range from 10 GeV to 2 TeV. From these sources, TeV candidates can be selected based on their GeV fluxes and $\gamma$-ray spectral indices. When focusing on TeV blazars, HSP blazars, especially HSP BL Lac objects (HBLs), can serve as a good data set from which to select TeV candidates. Several HSP blazar catalogs, such as the second Wide-field Infrared Survey Explorer (WISE) High Synchrotron Peak Catalog (2WHSP, \citealt{2017A&A...598A..17C}) and the catalog of extreme and high-synchrotron peak blazars (3HSP; \citealt{2019A&A...632A..77C}) have been utilized in the search for blazars that are likely to be detected by their TeV energies. Using artificial neural network (ANN) machine-learning (ML) algorithms, \cite{2019ApJ...887..104C} searched for HBL candidates in the collections of unclassified blazars and unassociated sources observed with Fermi-LAT, relying on the GeV data obtained from the 3FGL. They then analyzed the Fermi spectra of the HBL candidates and used an EBL model to identify TeV candidates. In other bands outside of the GeV $\gamma$-ray band, TeV blazars also show unique characteristics. \cite{2013ApJS..207...16M} have suggested that TeV BL Lac objects have a higher X-ray to infrared flux ratio and a lower WISE magnitudes and that these characteristics could be used to find TeV candidates. In an alternative approach, \cite{2020MNRAS.491.2771C} combined X-ray, $\gamma$-ray, and infrared data to analyze the clustering of TeV BL Lac objects in the entire BL Lac objects and managed to identify some TeV candidates.

Recently, the Fermi-LAT collaboration released the fourth Fermi-LAT Gamma-ray source catalog (4FGL-DR1, eight years of exposure; \citealt{2020ApJS..247...33A}), a fourth catalog of AGNs detected by the Fermi-LAT (4LAC, \citealt{2020ApJ...892..105A}), and a second release of 4FGL data (4FGL-DR2, 10 years of exposure; \citealt{2020arXiv200511208B}). 4FGL contains more gamma-ray sources than the previous source catalog, has higher energy upper-limits, and has longer exposure times \citep{2015ApJS..218...23A,2020ApJS..247...33A}. In addition, the reports provide wide cross-matching with radio observations (VLBI Radio Fundamental Catalog, e.g., RFC; see \citealt{2019MNRAS.485...88P}) and optical observations (Gaia Data Release 2, e.g., Gaia-DR2, see \citealt{2018yCat.1345....0G}) . These advances facilitate the exploration for candidates for TeV observations. In this paper, we present a new approach to search for TeV candidates in the 4LAC HBLs, which incorporates two steps: 1) radio, optical, and GeV ${\gamma}$-ray data are combined and several ML algorithms are employed to search for possible TeV candidates (PTCs) in the 4LAC HBLs; 2) the ${\gamma}$-ray spectra of the PTCs are analyzed by using the longer exposure times and higher energy upperlimits of the new Fermi data than reported in the published catalogs. The EBL model, FOV, and sensitivity of CTA and LHAASO are then used to evaluate the likelihood of a PTC being detected.

The remainder of this paper is organized as follows. In Section 2, we describe how ML was used to select PTCs from the 4LAC HBLs. In Section 3, we briefly review the Fermi spectral analysis of the PTCs and the EBL correction. In Section 4, we compare the corrected spectra with the sensitivities of the CTA and LHAASO. A discussion of the results and our conclusions are given in Section 5. Throughout the paper, we assume a Hubble constant $H_{0} = 75\rm \ km\ s^{-1}\ Mpc^{-1}$, the matter energy density $\Omega_M = 0.27$, the radiation energy density $\Omega_r = 0$, and the dimensionless cosmological constant $\Omega_{\Lambda} = 0.73$. We set the random seed as ``120'' in the algorithms with random processes.

\section{PTC SELECTION USING ML} \label{sec:classification}

ML techniques are popular in the field of astronomical data mining and data analysis \citep{2010IJMPD..19.1049B, 2012MNRAS.424L..64M, 2016MNRAS.462.3180C, 2016ApJ...820....8S, 2017MNRAS.470.1291S, 2019arXiv190407248B, 2019ApJ...872..189K, 2019ApJ...887..134K,2020MNRAS.498.1750A, 2021MNRAS.505.1268F, 2021RAA....21...15Z}. According to whether the classification of a sample is given, ML can be mainly divided into supervised machine-learning (SML) and unsupervised machine-learning (USML) techniques \citep{2019arXiv190407248B}. SML approaches are usually used for classification and regression. USML approaches focus on clustering and dimensionality reduction by searching for potential relationships among the given samples. Numerous algorithms are applied to SML classification, including logistic regression (LR), decision trees, random forest(RF), support vector machines (SVM), NNs, and Bayesian networks (see, e.g.,  \citealt{2012msma.book.....F,Kabacoff2015R} etc.). In SML classification, the data set contains a certain number of objects, with each object containing several features and a label \citep{2020arXiv200511208B}. Features are usually observations that characterize the physical properties of the object, while the labels mark the classes.  A classification algorithm model learns the relation between the features of known objects and their labels. The model can then be employed to evaluate a potential class of objects without clear classifications based on their features. The simple application of an SML classifier involves several steps, including data set preparation, model training, model optimization, model testing, and finally prediction.

Generally, for the binary classification of an unknown object between class A and class B, SML classifiers compute likelihood possibilities, $L_{\rm A}$ and $L_{\rm B}$, rather than give a classification directly. $L_{\rm A}$ or $L_{\rm B}$ correspond to the probability that the object belongs to A or B, respectively. Their relation can be expressed as $L_{\rm A} = 1 - L_{\rm B}$ \citep{2019ApJ...887..104C}. A binary ``soft'' classification can be achieved by selecting an $L_{\rm th}$ classification threshold in the range of $0 \sim 1$. In standard SML binary classification, the boundary between class A and class B is clear and, after training, the classifier is always applied to an independent unknown data set. In the context of this paper, the two types of objects are HBLs detected in the TeV energy band (labeled as TeV) and HBLs which are not detected in the TeV energy band (labeled as non-TeV). However, if an object is labeled as ``non-TeV'', this does not mean that it does not emit TeV $\gamma$-rays; it may not have been detected simply because of  observational limitations. So, there will be TeV sources in the non-TeV data set. With soft SML classifiers, it is possible to train the model on these kinds of data sets to establish a probability space. In the probability space, we can search for TeV candidates among the non-TeVs that bear some resemblance to a TeV source. This makes it possible to train, optimize, and test an SML model in the same way as the standard process, so as to obtain a result with a higher level of confidence. The SML model can then be employed to compute the likelihood probabilities for each source to create a probability space, after which, the PTCs can be selected.

When evaluating a classifier, there are different metrics for performance evaluation that can be used, such as accuracy, recall, precision, and the receiver operating characteristic (ROC) curve \citep{2019arXiv190407248B}. Accuracy is widely used, but simple classification accuracy is usually dominated by large samples from imbalanced data sets \citep{2021RAA....21...15Z}. In this study, we used the balanced-accuracy (i.e., the average accuracy of positive and negative samples) rather than simple accuracy to evaluate the classifier performance.

Combining the results of multiple SML classification algorithms can reduce misclassifications \citep{2021RAA....21...15Z}. In our approach, we combined two SML methods, SVM \citep{vapnik1999nature} and LR (e.g., \citealt{Cox1958LR,walker1967LR}). These are popular SML classifiers used in astronomy. Scikit-learn, originally known as sklearn \citep{scikit-learn}, is a complete ML integration Python package that makes it easy to construct various ML algorithm models. We created the SVM and LR models using the Scikit-learn package in a Python environment (version 3.8.5). The SML classification flowchart used in the context is shown in the Table \ref{Tab1}.

\begin{table}
\centering
\caption{ML classification flowchart}\label{Tab1}
\begin{tabular}{l}
\hline \hline
\textbf{ML algorithm}: SVM and LR$\ \ \ $\textbf{data set}: 180 HBLs with 34 features\\
\hline
Stage 1: data set preparation\\
$\ \ \ $1. Sample selection \\
$\ \ \ $2. Cross-matching multiband data \\
Stage 2: model creation\\
$\ \ \ $1. SVM: \emph{sklearn.svm.SVC()}\\
$\ \ \ $2. LR: \emph{sklearn.linear-model.LogisticRegression()}\\
Stage 3: model optimization\\
$\ \ \ $1. Feature selectiong with RFE: \emph{sklearn.RFECV()} \\
$\ \ \ $2. Hyper-parameter combination grid search: \emph{sklearn.model$\_$selection.GridSearchCV()}\\
Stage 4: model training and test\\
Stage 5: model results\\
$\ \ \ $1. Single algorithm classification result \\
$\ \ \ $2. Combining classification results of different algorithms\\
\hline
\end{tabular}\\
\end{table}

\subsection{Data set preparation}

The starting point was 4LAC \citep{2020ApJ...892..105A}. For the Fermi-LAT observations made between 2008 and 2016 in the energy range from 50 MeV to 1 TeV, the 4LAC reports 2863 AGNs located at high galactic latitudes ($\left|{b}\right| \textgreater 10^{\circ}$)\footnote{List in $table\_4LAC\_h.fits$}. These AGNs include 2779 blazars, accounting for 97.8$\%$ of the total. There are 655 FSRQs, 1067 BL Lac objects, and 1077 blazars of an unknown type (BCUs).

By cross-matching 4LAC with 4FGL-DR2, Gaia-DR2, and RFC, we selected samples according to the following principles: (i) HBL; (ii) an analysis flag value of ``0'' in the Fermi-DR2 to ensure the $\gamma$-ray data had a high confidence level; (iii) a source with counterparts in the RFC and Gaia-DR2 catalogs; and (iv) a source redshift that is known to be suitable for subsequent EBL correction. We ended up with 180 HBLs in our data set, including 39 TeVs and 141 non-TeVs.

For each HBL in the data set, 21 parameters were directly collected from the multiple catalogs (see Table 2, columns 2-5). There are eight direct observations in 4LAC: [$\rm Signif$-$\rm Avg$] $-$ the significance of the source in $\sigma$ units over the 50 MeV to 1 TeV band; [$F_{\rm 1000}$] $-$ the integral photon flux from 1 to 100 GeV; [$E_{\rm 100}$] $-$ the energy flux from 100 MeV to 100 GeV; [$\gamma \ $-$ \rm{spectrum}\;\rm{type}$] $-$ the band spectral type; [$\alpha_{\gamma}$] $-$ the photon spectrum of $\gamma$ band; [$\rm Redshift$] $-$ the redshift of each source; [$E_{\rm pivot}$] $-$ the pivot energy at which the error for the differential flux is at a minimum; [$\nu_{\rm syn}$] $-$ the synchrotron peak frequency in the observer frame; [${\nu F_{\nu}}_{\rm syn}$] the energy flux at the synchrotron peak frequency. From 4FGL-DR2, we obtained [${\nu F_{\nu}}_{\gamma 1}$-${\nu F_{\nu}}_{\gamma 7}$], which is the mean energy flux over seven $\gamma$-ray bands: 50-100 MeV; 100-300 MeV; 300 MeV-1 GeV; 1-3 GeV; 3-10 GeV; 10-30 GeV; and 30-300 GeV. Gaia-DR2 provided [$mag_{\rm G}$], which is the mean magnitude in the G band (330 $\thicksim$ 1050 nm); [$mag_{\rm BP}$], which is the integral mean magnitude over the BP band (330 $\thicksim$ 680 nm); and [$mag_{\rm RP}$], which is the integrated mean magnitude over the RP band (630 $\thicksim$ 1050 nm). RFC provided the radio observations of compact extragalactic sources. The source catalog, \emph{astrometric global solution rfc-2020a}\footnote{\url{http://astrogeo.org/vlbi/solutions/rfc_2020a/}}, reported the available VLBI observations for the S, C, X, U, and K bands from 1980 to 2020. However, once the missing data was removed, only [$F_{\rm r}$], the radio flux in the X band (8.198 $\thicksim$ 8.950 GHz), was available. For the above parameters, we found the logarithm of the higher scale features (e.g., flux, energy, and peak frequency) to reduce the computational load during the subsequent steps.

We also defined some induced parameters. The energy flux of a single band cannot characterize the evolution of the $\gamma$-ray spectrum. Therefore, following \cite{2012ApJ...753...83A}, we therefore calculated the hardness ratios based on the SEDs of seven $\gamma$-ray bands, using:
\begin{equation}\label{eq1}
 hr_{\rm ij}=\frac {{\nu F_{\nu}}_{\gamma \rm j}-{\nu F_{\nu}}_{\gamma \rm i}}{{\nu F_{\nu}}_{\gamma \rm j}+{\nu F_{\nu}}_{\gamma \rm i}}
\end{equation}
where ${\nu F_{\nu}}_{\gamma \rm i}$ and ${\nu F_{\nu}}_{\gamma \rm j}$ are the SEDs corresponding to the seven Fermi bands, in which $\rm j = i + 1$. The value of $hr_{\rm ij}$ is always between -1 and 1 and it describes the hardness of the spectrum over the {\rm i} and {\rm j} bands. We also defined the spectrum unevenness parameter, $H_{\rm ijk}$, as follows:
\begin{equation}\label{eq2}
 H_{\rm ijk}=hr_{\rm ij} - hr_{\rm jk}
\end{equation}
where $H_{\rm ijk}$  is the parameter that characterizes the unevenness of the spectrum over the {\rm i}, {\rm j}, and {\rm k} bands. So, there are six hardness ratios and five spectral unevenness parameters in a total of seven Fermi bands. We then defined the flux ratio of the radio, optical, and $\gamma$-ray bands as:
\begin{equation}\label{eq5}
 \Phi_{\rm AB}=\rm log\ \frac { F_{\rm A}}{F_{\rm B}}
\end{equation}
where, $F_{\rm A}$ and $F_{\rm B}$ are the flux values for the radio flux, $F_{\rm r}$, optical flux, $F_{\rm o}$, and $\gamma$-ray flux, $F_{\gamma}$ (all in unit of $\rm erg \cdot cm^{-2} \cdot s^{-1} \cdot Hz^{-1}$). Based on the mean magnitude given by GAIA-DR2, we obtained the optical flux, $F_{\rm o}$, with a zero magnitude flux correction by using the following expression:
\begin{equation}\label{eq3}
 F_{\rm o}=F_{\rm G0} \cdot 10^{- \frac{mag_{\rm G}}{2.5}} \ \scriptstyle{[\rm mJy]}
\end{equation}
where $F_{\rm G0}=3.06 \times 10^{6}\ \rm mJy$ is the zero-point magnitude flux at 640 nm (i.e., near the center of the G band; \citealt{1990A&AS...83..183M}). Assuming that the $\gamma$-ray spectra conformed to a powerlaw (PL), we were able to calculate the $\gamma$-ray flux, $F_{\gamma}$, at $E_{\gamma} = 5\ \rm GeV$, by means of the following equation (e.g., \citealt{2014Ap&SS.352..819Y}):
\begin{equation}\label{eq4}
 F_{\gamma}=h\cdot \frac {F_{\rm integral(E_1 \sim E_2)} (1-\alpha_{\gamma})}{E_{2}^{1-\alpha_{\gamma}} - E_{1}^{1-\alpha_{\gamma}}} \cdot E_{\gamma}^{1-\alpha_{\gamma}}\ \scriptstyle {\rm [erg \cdot cm^{-2} \cdot s^{-1} \cdot Hz^{-1}]}
\end{equation}
where $F_{\rm integral(E_1 \sim E_2)}$ represents the integrated photon flux over the energy ranges $E_{\rm 1}$ and $E_{\rm 2}$, $\alpha_{\gamma}$ is the photon spectrum index of the $\gamma$ band, and $h=6.63\times 10^{-27} \ \rm erg\ s$ is the Planck constant. The spectral index, $\alpha_{\gamma}$, and the integral photon flux, $F_{\rm integral(E_{\rm 1} \sim E_{\rm 2})}$, were directly obtained from the 4LAC table. 

In the above scenarios, we compile a ML data set contains 180 objects with 35 features.

\begin{table}
\centering
\caption{The observations from the multiple source catalogs}\label{Tab2}
\begin{tabular}{cccccc}
\hline \hline
&\multicolumn{4}{c}{Direct Parameter}&Induced Parameter\\
\cline{2-5}
Catalog&4LAC&4FGL-DR2&Gaia-DR2&RFC&\\
\hline
\multirow{3}{*}{\rm Parameters}&$Signif$-$Avg$, $\rm log(F_{\rm 1000})$, $\rm log(E_{\rm 100})$&&$mag_{\rm G}$&&$hr_{\rm ij}$\\
&$V_{\rm index}$, $\gamma$-$spectrumtype$, $\alpha_{\gamma}$, $\rm Redshift$&$\rm log({\nu F_{\nu}}_{\gamma 1})$-$\rm log({\nu F_{\nu}}_{\gamma 7})$&$mag_{\rm BP}$&$F_{\rm r}$&$H_{\rm ijk}$\\
&$E_{\rm pivot}$, $\rm log(\nu_{\rm syn})$, $\rm log({\nu F_{\nu}}_{\rm syn})$&&$mag_{\rm RP}$&&$\Phi_{\rm ro}$, $\Phi_{r\gamma}$, $\Phi_{o\gamma}$\\
\hline
\end{tabular}\\
{Note: The observations obtained from 4LAC, 4FGL-DR2, Gaia-DR2 and RFC, etc. The last column is the induced parameter calculated with the multiple observations.}
\end{table}

\subsection{\rm ML classification model}

Based on the Scikit-learn package \citep{scikit-learn}, the SVM classifier was built with \emph{sklearn.svm.SVC()}, while the LR model was established using \emph{sklearn.linear-model.LogisticRegression()}. In Section 2.1, we compiled a ML data set containing 180 pbjects with 35 features. Using the data set , we trained, optimized, and tested the classifiers in turn. ML data sets are usually divided into training, validation and test sets. Since the data set used here only contained 180 HBLs, it was too small to be further divided. We adopted a 5-fold cross-validation data set division method available within the Ssikit-learn package. This divided the data set equally into five ``fold''; four of these were used to train the models and the fifth was reserved for testing them. This was repeated five times. In this way, we obtained a mean value for the classifier performance across five iterations of training and testing. Cross-validation is an effective way to avoid the increasing randomness and over-fitting that can result from having an insufficient number of samples. 5-fold cross-validation is used both in the model optimization of feature selection and hyper-parameter combined searching. 

\begin{figure}
\centering
\includegraphics[height=6cm,width=8cm]{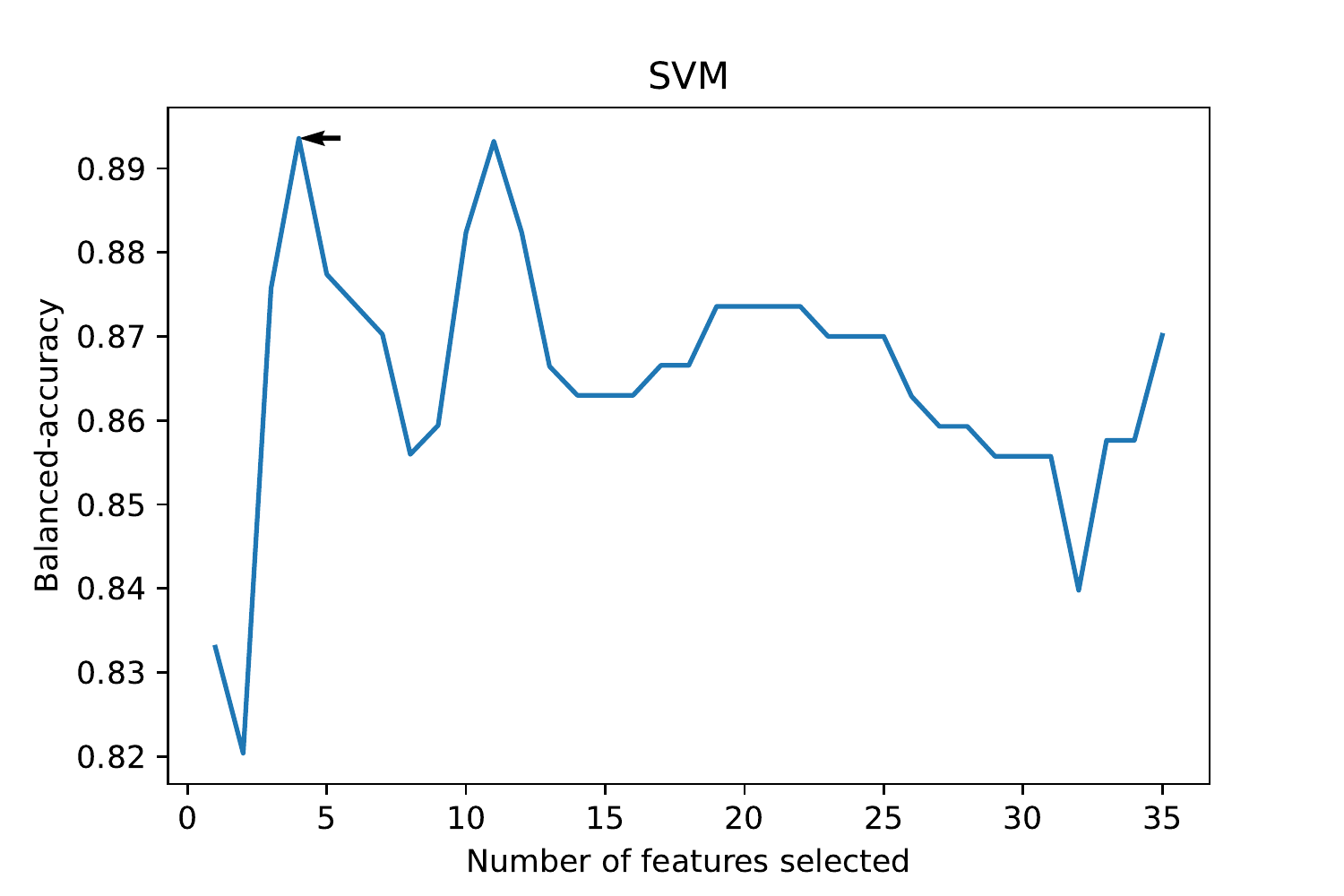}
 \includegraphics[height=6cm,width=8cm]{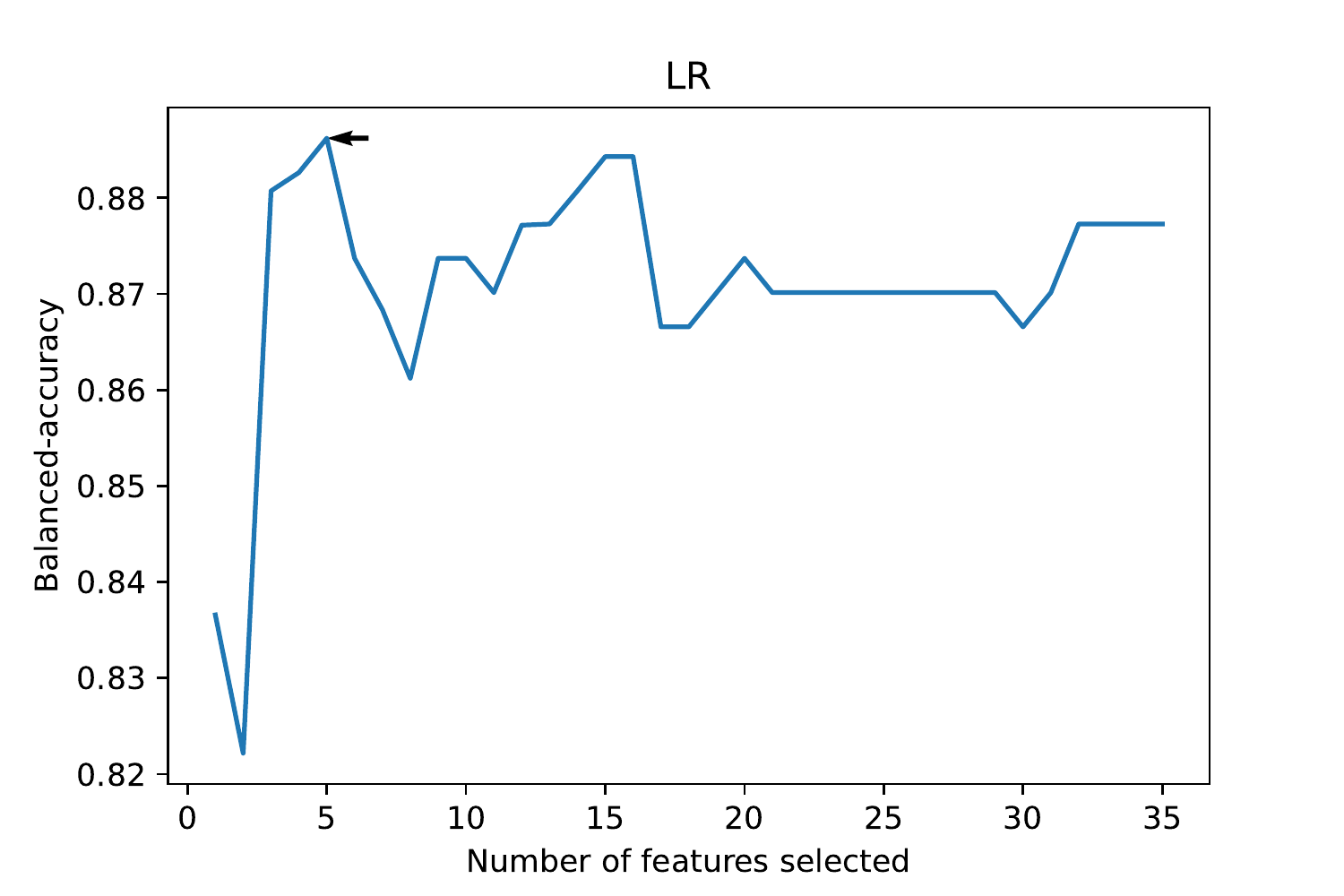}
\caption{The RFE curve of two classifiers. The left panel shows the relationship between balanced-accuracy and the number of features in REF of SVM, while the right panel shows the relationship in RFE of LR. The black arrow marks the location of the peak}
 \label{fig1}
\end{figure}

\begin{figure}
\centering
\includegraphics[height=8.5cm,width=8.5cm]{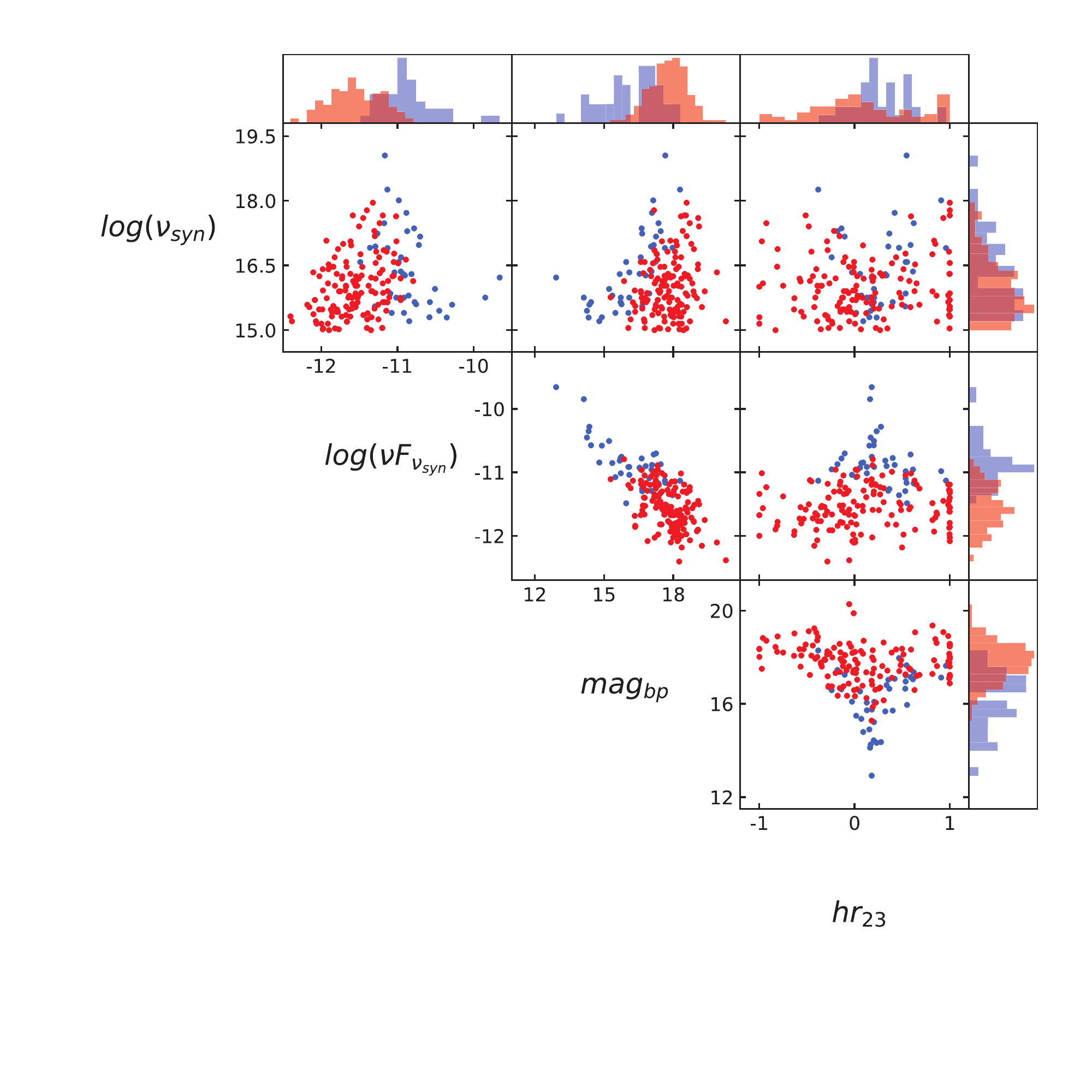}
 \includegraphics[height=8.5cm,width=8.5cm]{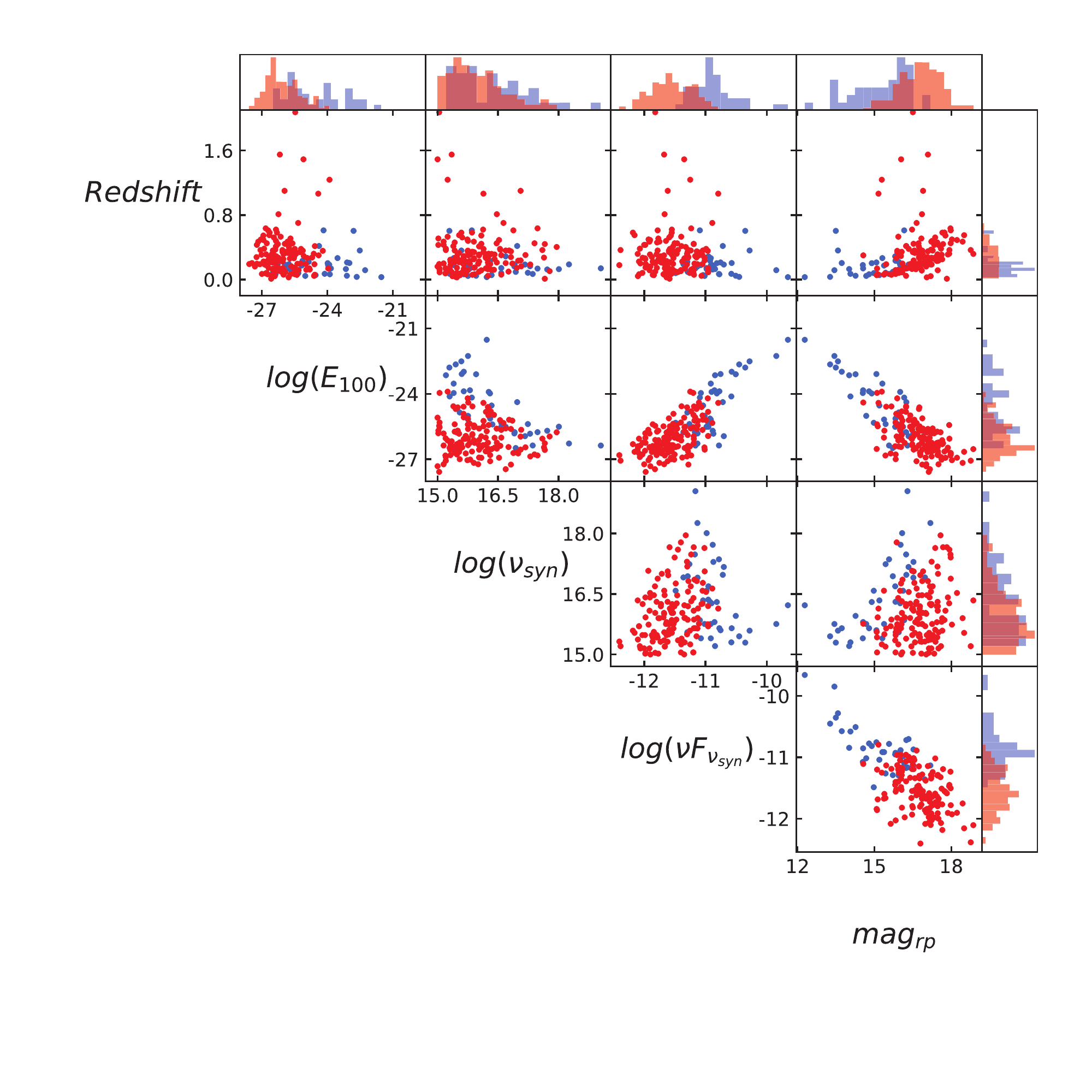}

\caption{Scatter plot of TeV and non-TeV in the parameter space constructed by the features obtained from RFE. The left panel shows the SVM parameter space, while the right panel represents the LR parameter space. The blue symbols represent TeVs and the red symbols represent non-TeVs. The outer part of each panel is the normalized distribution of each parameter.}
 \label{fig2}
\end{figure}

Each object in our ML data set contained 35 features; however, because they could have a direct and significant influence on the classification results, not all of the features were suitable for both classifiers. Either the two sample test \citep{2019ApJ...872..189K, 2019ApJ...887..134K, 2021RAA....21...15Z} or dimensionality reduction is often used for ML feature selection. Whereas, the Scikit-learn package provides a recursive feature elimination (RFE) approach. RFE selects features by recursively considering smaller and smaller sets of them in specific ML models to improve performance. We therefore adopted the RFE approach with 5-fold cross-validation. 
Different ML models contain inner parameters that affect the performance of the model. These are called hyper-parameters in SML. So, SVM contains three hyper-parameters, including the \emph{kernel}, kernel coefficient \emph{gamma} (not in linear kernel), and regularization parameter \emph{C}. When selecting features with RFE in SVM, we fixed the hyper-parameter of SVM: linear kernel with a regularization parameter $C=1.0$. Faced with an unbalanced data set, the parameter \emph{class$\_$weight} was set to ``balanced''. 
The LR classifier requires the hyper-parameters of \emph{solver} and regularization parameter \emph{C}. 
When selecting features with RFE in LR, we fixed the hyper-parameter of LR, \emph{lbfgs} kernel with regularization parameter $C=1.0$. The parameter \emph{class$\_$weight} is set to ``balanced'', and the maximum number of iterations \emph{max$\_$iter} is set to 500. The results of feature selection are shown in Table \ref{Tab3}, and the RFE curves (balanced-accuracy versus number of features) of the two classifiers are shown in Figure \ref{fig1}. Four features were used in the SVM classifier to obtain the largest balanced-accuracy (see the left panel in Figure \ref{fig1}), while only five features were suitable for the LR classifier (see the right panel in Figure \ref{fig1}). The distributions of the two sample labels in the parameter space constructed by the features obtained by the REF approach are shown in Figure \ref{fig2}. The left panel represents the parameter space for SVM RFE, while the right panel shows the parameter space for LR RFE. In Figure \ref{fig2}, there are different clustering characteristics between the TeVs (blue symbols) and non-TeVs (red symbols). For example, the TeVs exhibit a higher synchrotron peak SED and GeV $\gamma$-ray integrated energy fluxes, but lower optical magnitudes and redshifts. The different cluster distributions indicate that the TeV HBL sources can be roughly separated from the entire HBL population by using multiwavelength data. Consequently, adopting multiwavelength data to train the SML model was effective.

Next, we searched for hyper-parameter combinations corresponding to the highest balanced-accuracy. Using the features selected above for the two classifiers, we adopted a hyper-parameter combination grid search method with 5-fold cross-validation from Scikit-learn. The results are shown in Table \ref{Tab3}. The \emph{linear} kernel SVM algorithm with a regularization parameter, \emph{C}, of 100 performed well, giving a higher testing balanced-accuracy of 0.895 and the training  balanced-accuracy of 0.908. For the LR classifier with the \emph{lbfgs} solver, a regularization parameter, \emph{C}, of 0.1 achieved a higher training and testing balanced-accuracy of 0.904. There was no obvious over-fitting by the two classifiers. The SVM and LR models were then trained on the whole data set and we computed the likelihood of possible $L_{\rm TeV}$ values for each HBL.

\begin{table}[h]
\centering
\caption{The optimization results of classifiers}\label{Tab3}
\begin{tabular}{c|cccc}
\hline \hline
Classifier&Features&Hyper-parameter&Training Balanced-accuracy&Testing Balanced-accuracy\\
\hline
\multirow{2}{*}{SVM}&$\rm log(\nu_{\rm syn})$ , $\rm log({\nu F_{\nu}}_{\rm syn})$&Kernel: \emph{linear}&\multirow{2}{*}{0.908}&\multirow{2}{*}{0.895}\\

&$mag_{\rm BP}$, $hr_{\rm 23}$ &C: \emph{100}&&\\
\hline
\multirow{2}{*}{\rm LR}&$\rm {Redshift}$, $\rm log(E_{\rm 100})$ , $\rm log(\nu_{\rm syn})$ &\rm {Solver}: \emph{lbfgs}&\multirow{2}{*}{0.904}&\multirow{2}{*}{0.904}\\
&$\rm log({\nu F_{\nu}}_{\rm syn})$, $mag_{\rm RP}$&C : \emph{0.1}&&\\
\hline
\end{tabular}\\
{Note: Column 1: classifiers.
Column 2: the features obtained from RFE.
Column 3: the hyper-parameter combination corresponding to the highest balanced-accuracy obtained from GridSearch.
Column 4-5: the balanced-accuracy of the training and test set of the different classifiers in the cross-validation.
}
\end{table}

\subsection{\rm ML classification results}

\begin{figure}[h]
\centering
\includegraphics[height=6.5cm,width=9cm]{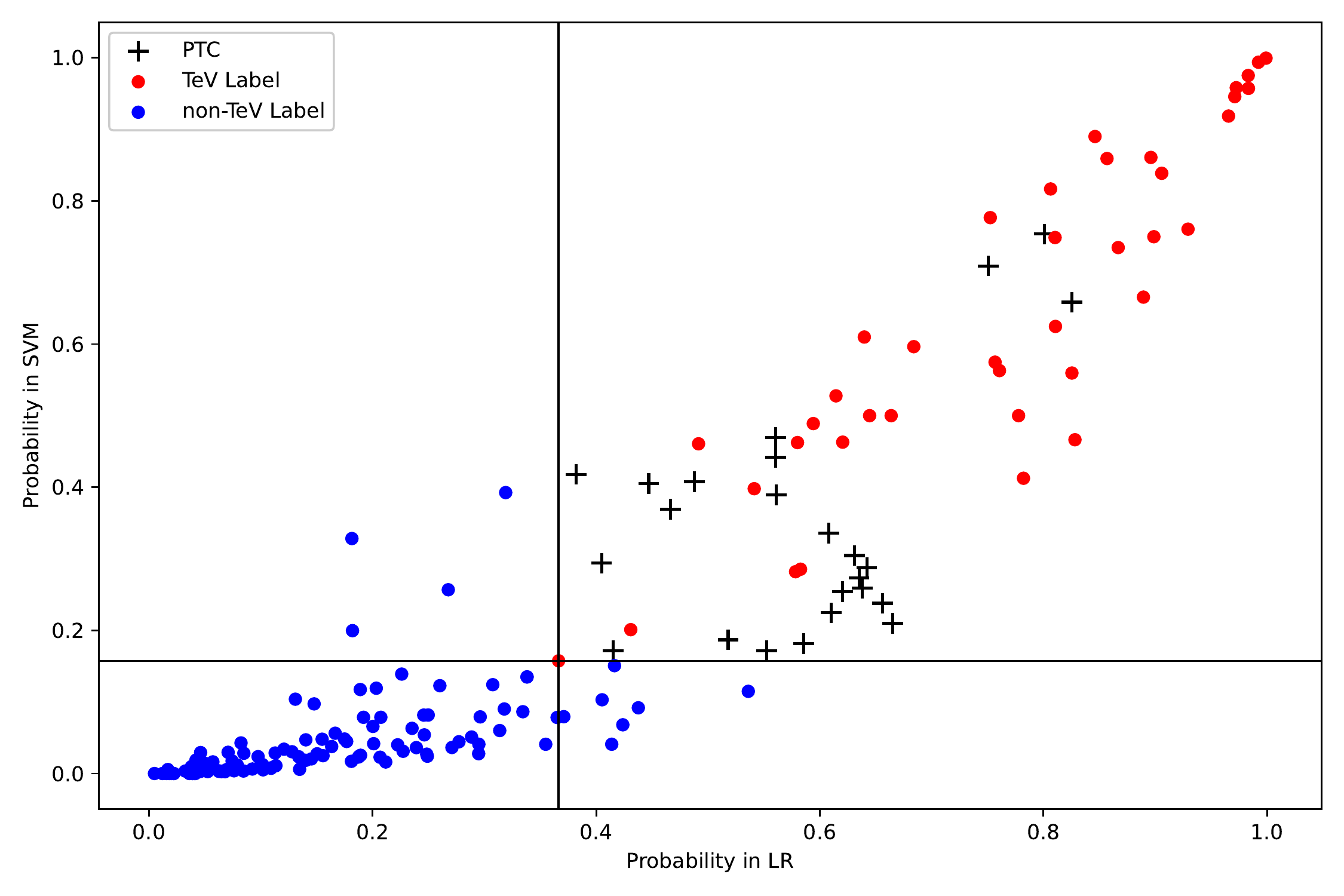}
 \caption{The distribution of 180 HBLs in probability space. The red symbols represent the TeVs, the blue symbols represent non-TeVs and the black symbols represent the PTCs obtained with our method.}
 \label{fig3}
\end{figure}

We built an $L_{\rm TeV}$ probability space in the SVM and RF models (see Figure \ref{fig3}). The red symbols represent the TeVs and the blue symbols represent non-TeVs. Taking into account the known TeV sources with the lowest probabilities predicted by the two classifiers (see the black lines in Figure \ref{fig3}), 24 PTCs (see the black symbols in Figure \ref{fig3}) were detected by our ML classifiers. The detailed information of the 24 PTCs is listed in Table \ref{Tab4}. Misclassifications are inevitable in ML, so, whether PTCs can be effectively detected by a TeV detector in this way requires further discussion.

\section{Fermi spectral analysis and EBL correction} \label{sec:Fermi}

The newest release of the Fermi $\gamma$-ray source catalog, 4FGL-DR2, contains GeV $\gamma$-ray spectral data for a 10 year period (2008-2018) in seven energy bins in the energy range of 50 MeV-500 GeV. By turning to more than 12 years of $\gamma$-ray observations provided by the Fermi Science Support Center (FSSC), we were able to analyze the $\gamma$-ray spectra of the PTCs obtained using the approach described above by utilizing more Fermi data than provided in the published catalogs. As the Fermi data were updated to P8R3 on 2008 November 12, we used the Fermi P8R3 data from 2008 October 1 to 2020 October 1 (mission elapsed time, MET, from 244548001 to 623253605). The photon events in the energy range from 100 MeV to 1 TeV were selected using the default data quality and a $90^\circ$ zenith angle. We used the corresponding instrument response functions for \emph{P8R3 SOURCE V3}, the galactic interstellar emission model, gll$\_$iem$\_$v07 (i.e.,  gll$\_$iem$\_$v07.fits\footnote{\url{https://fermi.gsfc.nasa.gov/ssc/data/analysis/software/aux/4fgl/Galactic_Diffuse_Emission_Model_for_the_4FGL_Catalog_Analysis.pdf}}), and the new isotropic spectral template (iso$\_$P8R3$\_$SOURCE$\_$V3$\_$v1.txt).  Together with the Fermi science tool, \emph{Fermitools} (version v11r5p3), the open-source Python package, \emph{Fermipy} \citep{2017ICRC...35..824W}, was used to calculate the SEDs of the PTCs.

\begin{figure}[h]
\centering
\includegraphics[height=20.5cm,width=18cm]{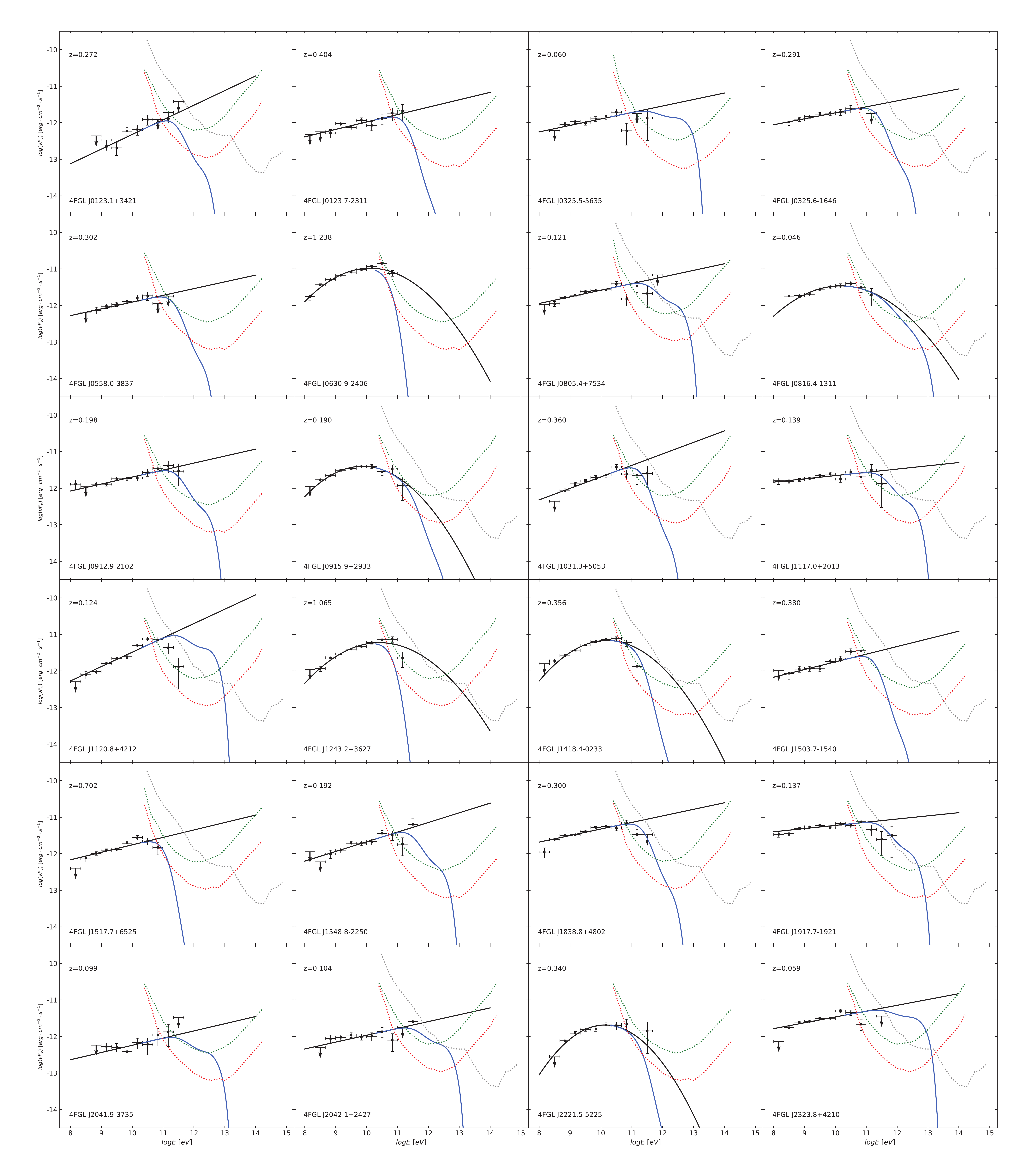}
 \caption{Twelve year averaged Fermi spectra of 24 PTCs. The black lines show the best-fitting lines found in the Fermi data analysis. The blue lines are the best-fitting lines with EBL correction. The black dotted line is the LHAASO sensitivity curve of $one\ year\ of\ operation$ \citep{2019arXiv190502773B}. The red and green point lines represent the CTA sensitivity curve of $\rm 50\ hrs\ 1yr^{-1}$ and $\rm 5\ hrs\ 1yr^{-1}$, respectively.}
 \label{fig4}
\end{figure}

Each spectra was divided into 12 energy bins. For each energy bin, we provided the energy flux, and there was a 1$\sigma$ error when $TS(Test Statistic) \geq 9$ ($\textgreater 3\sigma$), with the energy flux upper limit having a confidence level of 95$\%$ when $0\textless TS \leq 9$. We removed the energy bin when $ TS \leq 0$. Aside from the data points, \emph{Fermitools} provided the best-fitting lines for characterizing the evolution of the spectrum when analyzing the Fermi spectra. The 12 year average $\gamma$-ray spectra of the 24 PTCs are shown in Figure \ref{fig4}. Six sources are displayed as a LogParabola (LP) spectrum in the $\gamma$-ray band and the other 19 PTCs have PL-type SEDs.

TeV $\gamma$-rays crossing interstellar space are attenuated by $\gamma + \gamma \to e^+ + e^-$ through their interaction with EBL and CMB photons in wavelengths in the region of $0.1\sim 1000\ \mu \rm m$ \citep{1967PhRv..155.1408G,2005ApJ...618..657D}. Depending on the redshift, EBL absorption may only have a strong effect on the flux above a few tens of GeV. However, the best-fitting lines were mainly dominated by Fermi's low energy region, because there are higher photon counts in the low-energy region. The currently published Fermi source catalog has a detection energy upper limit of 1-3 TeV, although Fermi-LAT can detect higher energy photons. An increase in the detection energy band leads to a decrease in the effective detection area, an increase in the systemic uncertainty, and an insufficient high energy photon count. So, the Fermi spectra provide a good indication of the shape of the intrinsic source spectrum and can be extrapolated in the absorbed band.

We extended the Fermi best-fitting lines to the Fermi spectra to an energy of 100 TeV (see the black lines in Figure \ref{fig4}) assuming the spectra had an EBL optical optical depth, i.e., $\tau(E_\gamma,z)$, of 0. \cite{2008A&A...487..837F} and \cite{2017A&A...603A..34F} have provided an EBL optical depth table that takes into account the contribution of EBL photons. Using the EBL model from \cite{2008A&A...487..837F} and \cite{2017A&A...603A..34F}, we first calculated the EBL optical depth for all 24 TPCs from 20 GeV to 100 TeV, then corrected the Fermi spectra. The corrected best-fitting Fermi lines are shown in blue in Figure \ref{fig4}.

\section{Comparison of TeV flux and detection sensitivity }\label{sec:EBL correction}

The upcoming CTA and LHAASO will form the next generation of TeV detectors. They will be characterized by an extremely high sensitivities and large FOVs. The CTA is a new generation IACT, which contains two arrays. The North array is located at $28^{\circ}.7$, while the South array is at $-24^{\circ}.7$. The sensitivity of the southern array is slightly higher than that of the northern one. Drawing upon these two arrays, the CTA can achieve all sky observations in the energy range of 20 GeV-10 TeV. The sensitivity of the CTA is affected by the zenith angle ($z_{\rm max}$), location, and the Earth's magnetic field. Using the sensitivities provided by the CTA online calculator \footnote{The sensitivities of CTA are available at \url{https://www.cta-observatory.org/science/cta-performance/}}, for each PTC at a decl. of $b$, we obtained the CTA sensitivity curves for $\rm 5 hrs\ 1 yr^{-1}$ and $\rm 50 hrs\ 1 yr^{-1}$ as follows: [$b\geq 88^{\circ}.7$] $-$ sensitivity of the northern array at $z_{\rm max}=60^{\circ}$; [$58^{\circ}.7\leq b\textless 88^{\circ}.7$] $-$ sensitivity of the northern array at $z_{\rm max}=40^{\circ}$; [$2^{\circ}.7\leq b\textless 58^{\circ}.7$] $-$ sensitivity of the northern array at $z_{\rm max}=20^{\circ}$; [$-54^{\circ}.7\leq b\textless 2^{\circ}.7$] $-$ sensitivity of the southern array at $z_{\rm max}=20^{\circ}$; [$-84^{\circ}.7\leq b\textless -54^{\circ}.7$] $-$ sensitivity of the southern array at $z_{\rm max}=40^{\circ}$; [$ b\leq -84^{\circ}.7$] $-$ sensitivity of the southern array at $z_{\rm max}=60^{\circ}$. LHAASO, which is located in Daocheng, Sichuan province, China, aims to detect cosmic rays and $\gamma$-rays with an energy higher than 30 TeV using three detectors: KM2A, WCDA, and WFCTA. LHAASO can naturally survey half of the all sky from decl. $-20^{\circ}$ to $80^{\circ}$ in all weather when the zenith angle is set at $50^{\circ}$ \citep{2019ChA&A..43..457C}. For the PTCs in the LHAASO's FOV, we plotted the sensitivity curve for one year of operation  \citep{2019arXiv190502773B}, and the sensitivity curve is shown as a black dotted line in Figure \ref{fig4}. The CTA sensitivity curves for $\rm 50\ hrs\ 1 yr^{-1}$ and $\rm 5\ hrs\ 1yr^{-1}$ are shown by the red and green dotted lines, respectively.

We compared the results of the EBL-corrected PTCs with the CTA and LHAASO detection sensitivitied. There are 16 PTCs in the FOV of LHAASO, four of which are likely to be detected by LHAASO observations in light of the corrected Fermi spectrum. Out of the 24 PTCs, half are located in the northern sky area of the CTA northern array's FOV. The energy spectra of all PTCs are above the sensitivity of the CTA's $\rm 50\ hrs\ 1yr^{-1}$ observations, while only 13 PTCs are above the sensitivity of the CTA $\rm 5\ hrs\ 1yr^{-1}$ observations. Detailed information regarding the 24 PTCs is listed in Table \ref{Tab4}. The distribution of 24 PTCs in the FOV of CTA and LHAASO under the Galactic coordinate system is shown in Figure 5.

%\begin{landscape}
\begin{sidewaystable}[thp]
\centering
\caption{Information of the TeV candidates}\label{Tab4}
\begin{tabular}{cccccccccccc}
\hline
\multirow{2}{*}{Sourcename-4LAC}&\multirow{2}{*}{R.A.}&	\multirow{2}{*}{Decl.}&	\multirow{2}{*}{Redshift}&\multirow{2}{*}{$prob_{SVM}$}&	\multirow{2}{*}{$prob_{LR}$}&LHAASO&LHAASO&CTA-north&CTA-south&CTA&CTA\\
&&&&&&FOV&$one\ year\ operation$&FOV&FOV&$5\ hrs\ 1yr^{-1}$&$50\ hr\ 1yr^{-1}$\\
\hline\hline

4FGL J0123.1+3421	&	20.791	&	34.354	&	0.272	&	0.418	&	0.382	&	Y	&	N	&	Y	&	N	&	N	&	Y	\\
4FGL J0123.7-2311	&	20.938	&	-23.194	&	0.404	&	0.369	&	0.467	&	N	&		&	N	&	Y	&	N	&	Y	\\
4FGL J0325.5-5635	&	51.379	&	-56.591	&	0.06	&	0.171	&	0.552	&	N	&		&	N	&	Y	&	Y	&	Y	\\
4FGL J0325.6-1646	&	51.418	&	-16.781	&	0.291	&	0.171	&	0.415	&	Y	&	N	&	N	&	Y	&	Y	&	Y	\\
4FGL J0558.0-3837	&	89.523	&	-38.632	&	0.302	&	0.389	&	0.561	&	N	&		&	N	&	Y	&	Y	&	Y	\\
4FGL J0630.9-2406	&	97.741	&	-24.111	&	1.238	&	0.225	&	0.61	&	N	&		&	N	&	Y	&	N	&	Y	\\
4FGL J0805.4+7534	&	121.362	&	75.577	&	0.121	&	0.274	&	0.635	&	Y	&	Y	&	Y	&	N	&	Y	&	Y	\\
4FGL J0816.4-1311	&	124.112	&	-13.197	&	0.046	&	0.254	&	0.62	&	Y	&	N	&	N	&	Y	&	Y	&	Y	\\
4FGL J0912.9-2102	&	138.227	&	-21.045	&	0.198	&	0.238	&	0.656	&	N	&		&	N	&	Y	&	Y	&	Y	\\
4FGL J0915.9+2933	&	138.986	&	29.553	&	0.19	&	0.305	&	0.631	&	Y	&	N	&	Y	&	N	&	N	&	Y	\\
4FGL J1031.3+5053	&	157.845	&	50.884	&	0.36	&	0.47	&	0.56	&	Y	&	N	&	Y	&	N	&	Y	&	Y	\\
4FGL J1117.0+2013	&	169.271	&	20.229	&	0.139	&	0.182	&	0.586	&	Y	&	N	&	Y	&	N	&	Y	&	Y	\\
4FGL J1120.8+4212	&	170.201	&	42.204	&	0.124	&	0.336	&	0.608	&	Y	&	Y	&	Y	&	N	&	Y	&	Y	\\
4FGL J1243.2+3627	&	190.812	&	36.459	&	1.065	&	0.754	&	0.801	&	Y	&	N	&	Y	&	N	&	N	&	Y	\\
4FGL J1418.4-0233	&	214.606	&	-2.559	&	0.356	&	0.287	&	0.642	&	Y	&	N	&	N	&	Y	&	Y	&	Y	\\
4FGL J1503.7-1540	&	225.925	&	-15.683	&	0.38	&	0.442	&	0.561	&	Y	&	N	&	N	&	Y	&	Y	&	Y	\\
4FGL J1517.7+6525	&	229.436	&	65.424	&	0.702	&	0.408	&	0.488	&	Y	&	N	&	Y	&	N	&	N	&	Y	\\
4FGL J1548.8-2250	&	237.201	&	-22.847	&	0.192	&	0.187	&	0.518	&	N	&		&	N	&	Y	&	Y	&	Y	\\
4FGL J1838.8+4802	&	279.714	&	48.041	&	0.3	&	0.658	&	0.825	&	Y	&	N	&	Y	&	N	&	Y	&	Y	\\
4FGL J1917.7-1921	&	289.438	&	-19.363	&	0.137	&	0.21	&	0.665	&	Y	&	Y	&	N	&	Y	&	Y	&	Y	\\
4FGL J2041.9-3735	&	310.479	&	-37.587	&	0.099	&	0.294	&	0.405	&	N	&		&	N	&	Y	&	Y	&	Y	\\
4FGL J2042.1+2427	&	310.536	&	24.458	&	0.104	&	0.709	&	0.751	&	Y	&	N	&	Y	&	N	&	Y	&	Y	\\
4FGL J2221.5-5225	&	335.391	&	-52.431	&	0.34	&	0.405	&	0.447	&	N	&		&	N	&	Y	&	N	&	Y	\\
4FGL J2323.8+4210	&	350.975	&	42.183	&	0.059	&	0.259	&	0.638	&	Y	&	Y	&	Y	&	N	&	Y	&	Y	\\

\hline
\end{tabular}\\
{Note:Column 1-4: the information of the candidates we obtained. Column 5-6: the probabilities of candidates given by the different classifiers in SML. Column 7: a flag indicated whether the candidates can be detected by LHAASO. ``Y'' represents that the source is located in the LHAASO FOV, while the ``N'' represents that the source is out of the LHAASO FOV.}
\end{sidewaystable}
%\end{landscape}

\begin{figure}[h]
\centering
\includegraphics[height=6cm,width=8cm]{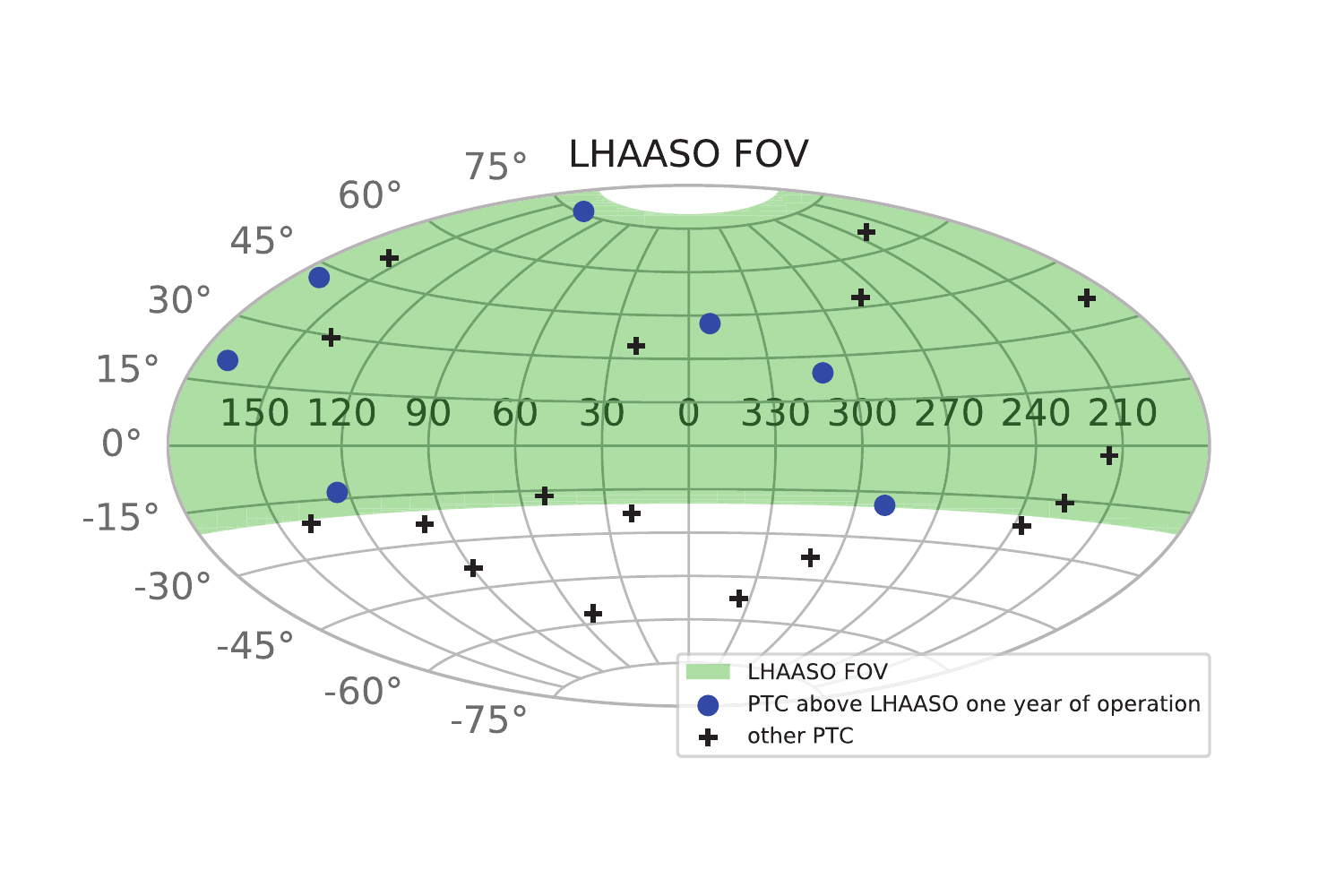}
 \includegraphics[height=6cm,width=8cm]{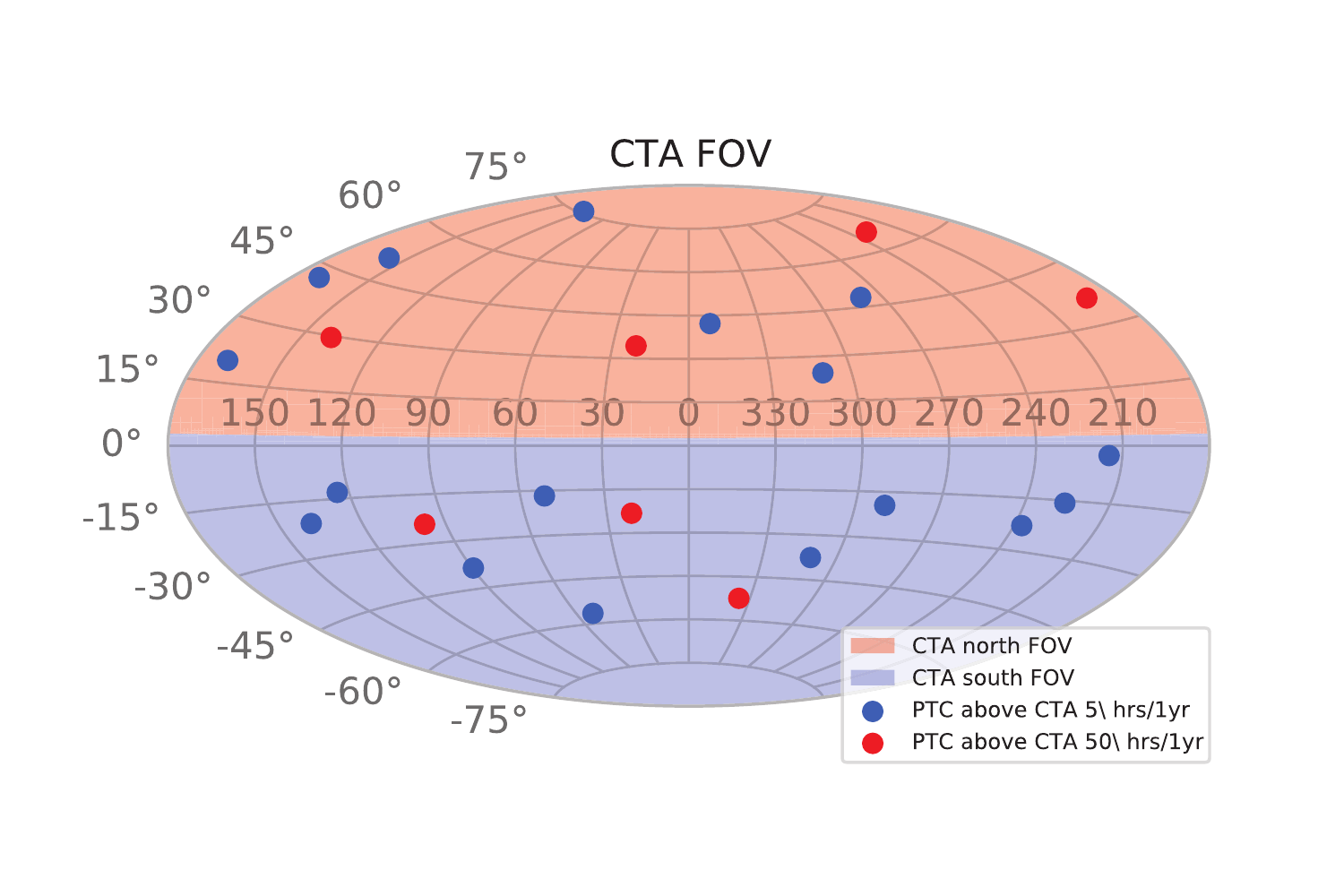}

\caption{All sky distribution of PTCs in J2000 coordinate. The left panel shows the FOV of LHAASO, and the right panel represents the FOV of CTA.}
 \label{fig5}
\end{figure}

\section{Conclusion and Discussion} \label{sec:Conclusion and Discussion}

In this study, we split the process of searching for TeV candidates in the 4LAC HBLs into two steps. First, we used SVM and RF algorithms to search for PTCs in the 4LAC HBLs by combining radio, optical, and GeV $\gamma$-ray data. This search revealed 24 PTCs that were above the minimum confidence standard in the SML probability space. We then analyzed PTC $\gamma$-ray spectra costructed from 12 years of Fermi observations and corrected them using an EBL model. Taking into account the sensitivity of the next generation CTA and LHAASO, we suggested four candidates for LHAASO observations and 24 candidates for CTA observations.

Current TeV detectors (e.g., IACTs and EAS arrays) are hindered by their limited sensitivities and FOVs, strong background interference, and their susceptibility to bad weather. The sensitivities of CTA and LHAASO are excepted to be approximately an order of magnitude higher than those of current detectors \citep{2013APh....43..171B,2014APh....54...86C}. CTA is excepted to yield good performance in the soft TeV band ($E\ \textless\ 10\ \rm TeV$), while LHAASO will focus on higher energy phenomena. The observation energy range of CTA and LHAASO may well therefore turn out to be complementary, enabling high detection sensitivity throughout the whole TeV energy range.

\begin{figure}[h]
\centering
\includegraphics[height=8cm,width=12cm]{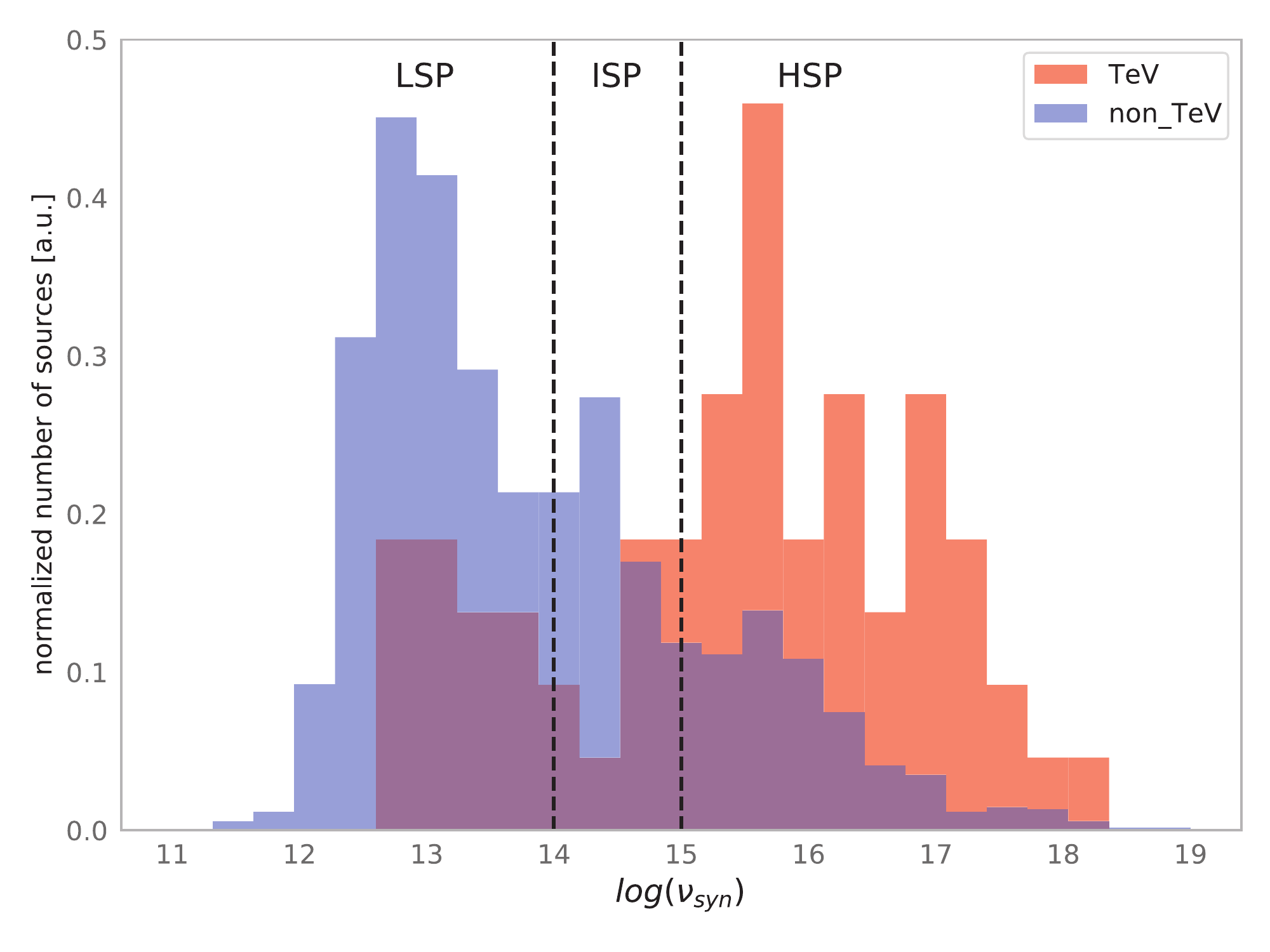}
 \caption{Normalized distribution histogram of the synchrotron radiation peaked frequency of TeV and non-TeV blazars in 4LAC. the red region represents the distribution of TeV blazars, and the blue region shows the distribution of non-TeV blazars.}
 \label{fig6}
\end{figure}

The samples used in the present work were limited to HBLs. However, there are also TeV FSRQs, TeV IBLs, and TeV LBLs in the overall population of TeV blazars. Emission models of blazars suggest that the peak frequencies of the two bumps in blazar SEDs are correlated \citep{2010ApJ...716...30A}. A higher peak frequency in a synchrotron spectrum usually means there is a higher peak frequency for the high energy bump. This is why TeV blazars are dominated by HBLs. For example, 43 4LAC TeV HBLs account for 68.3$\%$ of the 4LAC TeV blazars, while the 283 Fermi HBLs account for approximately just 10$\%$. It can be seen in Figure \ref{fig6} that the distributions of the TeVs and non-TeVs in terms of the peak frequency of synchrotron emissions are quite different. If sources displaying intermediate or low synchrotron peaks are accommodated within studies, a sample selection bias is inevitable (e.g., \citealt{2012ApJ...744..192R,2020MNRAS.492.5377L,2021RAA....21...15Z}). On the other hand, in the 4LAC, the BL Lac objects with peaked frequency of synchrotron emission $\nu_{\rm syn}\textgreater 10^{15} \rm{Hz}$ are always recognized as HBLs, it is hard to clearly distinguish EHBLs from HBLs. \cite{2020MNRAS.493.2438A} proposed that the search for TeV blazars may benefit from considering HSP and EHSP as a whole. It is consistent with our sample selection criteria.

The best-fitting lines after EBL correction for most PTCs were a better match for the Fermi data points than the uncorrected fits. This also confirms that the part of the BL spectra of blazars that break at the GeV level can be attributed to EBL attenuation \citep{2005ApJ...618..657D}. However, several sources did not match the best-fitting lines so well. For example, the spectra of 4FGL J0630.9-2406 (z=1.238) and \emph{4FGL J1243.2+3627} (z=1.065) showed an obvious cutoff after the EBL correction  was made that is not seen in the data points. This means that the result of the EBL correction-based TeV fluxes limitation for these two sources may be unacceptable. The excess of TeV $\gamma$-rays in high-redshift blazars remains an open issue. Several theories have been put forward to explain the spectra of hard $\gamma$-rays, such as axion-like particles \citep{2008PhRvD..77f3001S,2009PhRvD..79l3511S}, or Lorentz invariance violation \citep{2000PhLB..493....1P}, but, to date, there is no definitive conclusion. In addition, the spectra of PTCs \emph{4FGL J1120.8+4212} (z=0.124) and \emph{4FGL J2323.8+4210} (z=0.059) has earlier breaks than the best-fitting lines in the GeV energy band. This manifested as a mismatch between the data points and the fitted line. It is possible that these resulted from the intrinsic nature of the spectra rather than EBL absorption. \cite{2006ApJ...653.1089L} and \cite{2008ApJ...677..884L} have indicated that the GeV break of blazars may result from the absorption in the board-line region. That thee break points of the PTC spectra we uncovered below 100 GeV provided evidence that the GeV breaks of a blazars do not have any single EBL origin.

The $\gamma$-ray spectra for some of the PTCs, such as \emph{4FGL J0325.5-5635}, \emph{4FGL J1548.8-2250}, and \emph{4FGL J2042.1+2427}, suggested the presence of a hard-soft-hard trend, which may correspond to a third component in addition to synchrotron radiation and inverse Compton scattering. The origin of the TeV peak remains another open issue. Exploring more HBLs with a TeV peak and investigating emission models, such as a secondary radiation model of the interactions between cosmic rays and galaxy background light in neighboring space, is planned for consideration in our future work.

\section*{Acknowledgements}
We thank the anonymous referee for their very constructive and helpful comments and suggestions, which greatly helped us to improve our paper. We particularly thank Dr. C. Y. Yang from the Yunnan Observatory who provided us with many helpful comments and suggestions. We also thank the Fermi collaboration and Gaia collaboration for their data support. This work is partially supported by the National Natural Science Foundation of China (Grant Nos. 11763005 and 11873043) and the Science and Technology Foundation of Guizhou Province (QKHJC[2019]1290). This work is also supported by the Graduate Research Foundation of Yunnan Normal University. We would like to express our gratitude to EditSprings (https://www.editsprings.com/) for the expert linguistic services provided.

%\bibliography{article}
%\bibliographystyle{aasjournal}

\end{document}